# SAR REDUCTION TECHNIQUES FOR WEARABLE APPLICATION

**Prithvi Raj**[*, 1] **and Ravi Prakash Dwivedi**[2]

**Abstract**—The biocompatible, noninvasive, lightweight, compact size and low manufacturing cost of the wearable antenna has attracted the researchers to work more in this field to bring the use of wearable antennas to the mainstream wearable devices industry. In recent years there has been great growth in the field of wearable electronics due to their unmatched functionalities and use cases, with the increased use of wearables there comes the concern about the safety in the usage of the antenna which directly points towards the SAR value of the particular antenna in use. This paper examines a wide range of antenna based on their SAR value and the techniques such as EBG, AMC, use of metamaterials, used to reduce the SAR for safe usage.

## 1. KEY TERMS

Wearable antenna, SAR, metamaterial, AMC, FSS.

## 2. INTRODUCTION

Specific Absorption Rate (SAR) is a crucial parameter used to examine the safety of wireless devices that emit electromagnetic fields, including antennas. When electromagnetic energy is radiated from an antenna, a portion of that energy is absorbed by the surrounding biological tissue, including the human body. The SAR value of an antenna is an important consideration when assessing the safety of wireless devices and making sure they adhere to legal requirements. SAR values are used to establish guidelines and limitations for exposure to electromagnetic radiation by calculating the energy absorbed by the body. To guarantee that wireless devices and antennas meet safety requirements and regulations, it is imperative for antenna designers and manufacturers to take SAR into account. An antenna's SAR is influenced by a number of variables, including antenna type, size, frequency range, power output, and proximity to the body. Understanding SAR is crucial for assuring commitment to safety norms as well as the safe usability of wireless devices. There are methods like use EBG structures, FSS structures, Metamaterials, compact antenna design, multi-band antenna design, are some of the ways used to reduce the back radiations of antennas to lower the SAR, some of which being more effective and promising are discussed later in this paper.

## 3. SAR (SPECIFIC ABSORPTION RATE)

The radiation absorbed in tissues by RF pulses is known as the specific absorption rate (SAR), It is expressed in watts per kilogram. Specific absorption rate (SAR) is one of the key factors to consider when determining the level of risk in the usage of wearable electronic devices. The safety in the usage of antenna is must to be examined before its real life on-body usage, the antenna should be non-intrusive, and have a low radiation toward human body and have a net SAR value lesser than 2 W/Kg as set by International Electronical Commission (IEC) [1]. Table 1, presents SAR values for different designs. Fig 1, shows a modal of wearable textile antenna on human body [2]. Each group uses a different set of tissue models and phantoms. The SAR values are affected by parameters relating to antenna and phantom separation, as well as varying phantom sizes and tissue compositions. [3], [4]. SAR values depend on variables including the type of antenna and its dimensions as well as the operating environment. Instead, these conclusions must be adjusted for the particular application. As can be observed in Table 1, Antennas operating at lower frequencies typically produce lower Specific Absorption Rate (SAR) values compared to higher frequency multiband antennas. However, multiband antennas that operate at higher frequencies often generate higher SAR values.

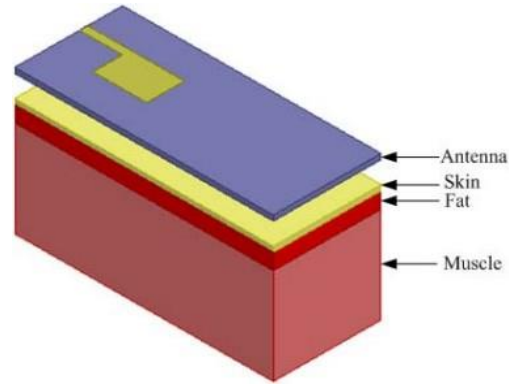

**Figure 1.** wearable textile antenna on human body

**Table 1.** SAR values of different antenna designs.

| source | Design | Dimensions (mm) | Frequency (GHz) | SAR 10gm (W/kg) |
|---|---|---|---|---|
| [3] | Rectangular patch antenna with inset feed | 51x45 | 2.45 | 0.486 |
| [4] | Slot inside rectangular patch | 70x70 | 2.45 | 0.23 |
| [5] | Two triangles with parallel slots | 70x50 | 1.198-4.055 | 0.0014 |
| [6] | Tuning fork shape antenna | 25x25 | 3.4 - 4.3, 4.7 - 8.4 | 1.919 |
| [7] | Rectangular patch antenna | 45x52 | 2.45 | 1.919 |
| [8] | Two stub and three slots patch antenna | 70x70 | 2.45 | 0.22 |
| [9] | C-shape etching slot | 18x19 | 3.09-3.94, 4.23-5.65 | 0.353 |
| [10] | Patch antenna with PIFA | 90x90 | 2.45 | 0.04 |
| [11] | Slotted rectangular patch | 70x60 | 2.45 | 0.043 |
| [12] | Pentagonal patch | 70x85 | 0.9,1.8 | 1.20,1.81 |
| [13] | L slot AMC on a semi-circular meta patch | 66.8x66.8 | 2.45 | 0.00544 |
| [14] | Metasurface antenna | 50x50 | 5 | 0.0975 |
| [15] | 2x2 EBG array | 34x34 | 5.8 | 0.094 |
| [16] | Rectangular patch | 120x120 | 2.45 | 0.00544 |

## 4. SAR REDUCTION TECHNIQUES

The use of wearable electronics for on body applications such as health monitoring, military applications, communication have increased in recent years and thus the safety in the usage of these wearables must be considered seriously and for this goal to achieve A number of techniques have been devised to keep SAR levels as low as feasible while preserving greater effectiveness. The antenna without a ground plane (i.e., a dipole) will have higher SAR values because the SAR of on-body antennas depends in part on near-field coupling to the body [17], [18], [19], [20]-[21]. As a result, many of the methods for decreasing the SAR of off-body emitting antennas focus on altering the ground plane [22]. SAR has a major influence on a wide range of applications, as demonstrated by this survey study. EBG, AMC structure is highly suggested for lowering SAR value. Ferrite sheets and metamaterials can be incorporated for further SAR reduction strategies.

## 4.1. A. SAR reduction using Electromagnetic Band Gap

The most appealing feature of EBG material is its stop-band and delayed response from the regular arrangement of the structure. The EBG structure features high impedance surfaces in a certain frequency range and is able to block surface wave propagation and adjusting radiation pattern. During on-body use, the EBG structure decreases the surface EM waves that propagate towards the human body. Also because EBG structure functions as a PMC, it improves antenna performance [23].The use of EBG was done to radiate back the radiations coming towards the human body [20]. Using EBG structural geometry at 2.4 GHz resulted in considerable changes in SAR, with the greatest SAR at 0.0545 W/kg and 5.41 W/Kg without EBG [24]-[28].Further Because the periodic dimension of EBG structureis generally approximately half-wavelength of the stopband, its huge size has limited its applicability in the low frequency range [29], [30].

## 4.2. SAR reduction using Metamaterials.

Metamaterials have attracted a lot of attention because of their unusual physical features and innovative applications. AMC structures typically have high electromagnetic surface impedance, which, in some range of frequencies, can prevent surface current transmission and act as a perfect magnetic conductor (PMC). In [31], author has made the use of artificial magnetic conductor (AMC) which has a slotted squared metal patch, a dielectric substrate along with a ground layer for reducing specific absorption rate (SAR) value of proposed PIFA antenna. AMC was fabricated on Taconic CRE-10 of dimension 54mm x 24mm x1.57mm with a 3 x 7 unit cell array. There was a reduction of 43.3 percent in theSAR after incorporating the AMC structure. In [32], design of dual band CPW fed antenna is proposed where a Gain improvement of 9.3 dB and 5.37 dB is observed along with the increase in the radiation efficiency by 48.4 percent and 35.7 percent at a frequency of 2.45 and 5.8 GHz respectively by the placement of metamaterial based structure having three layers consisting of a square metal ring around a square metal patch on the dielectric substrate, and then another metal layer as the ground plane below the substrate, this structure resulted in SAR reduction by 70 percent.

## 4.3. SAR reduction using Conductive materials.

A key consideration in antenna design is lowering specific absorption rate (SAR). Conducive materials can be used in the design as one method of achieving this. SAR can be decreased by the absorption or reflection of electromagnetic waves by conductive materials like metals and carbon-based compounds. Lower SAR values can be achieved by adding a conductive layer next to the antenna, which will deflect the energy emitted from the antenna away from the user's head or body. To further lessen the interaction of electromagnetic fields with adjacent objects and hence lower the SAR values, conductive coatings can be placed to the antenna. Yet, the inclusion of conductive materials can also have an impact on the antenna's overall functionality, changing both the radiation pattern and the gain. There are materials like conducting foam, conducting mesh, and conducting cloth. Wire-mesh shields and multilayered designs that can also be used for shielding the back radiations and reduce the SAR. Materials used as shields can be constructed of nickel, copper, or nickel-nickel.

## 4.4. SAR reduction using Frequency Selective Surface (FSS).

The distinctive stop-band or passband properties of frequency selective surfaces (FSS) are used to attempt to decrease SAR. In [33], to reduce the specific absorption rate (SAR) and increase the gain,a FSS textile-based superstrate of dimension (120 x 120 mm2) in the form of a square wave loop is backed by a textile-based compact monopole antenna (50.5 x 38 mm2). The square-shaped modified monopole radiator having rectangular slot defective ground structure radiates in the 2.45 GHz ISM with a bandwidth of 2.075 to 2.625 GH and for oblique incidence of signals up to 60°, achiral FSS offers a progressive increase in inductance and consistent overall performance for both the TE and TM polarization. The suggested antenna system with the FSS, with the sizes of 120x120x30 mm3, gives the max gain of 7.76 dB and the fractional bandwidth of 22.44 percent, the addition of FSS lowers back radiation, which in turn resulted in the reduction of SAR level by 95 percent.

## 4.5. SAR reduction using High impedance surface

The benefits of utilizing HIS include shielding the bodies from dangerous back radiation meanwhile maintaining the antenna's effectiveness, which may be impacted by the body's high conductivity. More-over, placing the antenna on the human body alone causes it to lose frequency but, when HIS is added, itbecomes effective and resilient to body stress and deformation. In [34], authors proposed the design of a compact (45x45x2.4 mm3) and strong antenna backed with a high-impedance surface (HIS) intended to function at 2.45 GHz for wearable applications. The integrated antenna with HIS performs admirably, with gains of 7.47 dB, efficiencies of 71.8 percent, and FBRs of 10.8 dB. The SAR is also lowered 95 percent.

**Table 2. Presents some of the widely used SAR reduction techniques.**

| source | Reduction method | Before Reduction(W/kg) | After Reduction(W/kg) | Operational Freq. (GHz) |
|--------|------------------|------------------------|-----------------------|-------------------------|
| [17]   | EBG              | 7.18 / 7.96            | 0.31 / 0.42           | 2.45/5.8                |
| [18]   | EBG              | 5.77/6.62              | 0.024 / 0.016         | 1.8/2.45                |
| [21]   | EBG              | 2.36                   | 1.77                  | 2.45                    |
| [23]   | EBG              | 2.814                  | 0.956                 | 2.54                    |
| [25]   | EBG              | 6.56                   | 0.0251                | 2.4                     |
| [34]   | HIS              | 7.51/8.64              | 0.0257/0.0358         | 2.45                    |
| [35]   | AMC              | 2                      | 0.29                  | 2.45                    |
| [36]   | AMC              | 4.2                    | 0.0257/0.0358         | 2.4                     |
| [37]   | Metamaterial     | 6.27                   | 0.0671                | UWB                     |
| [38]   | AMC              | 1.4                    | 0.7                   | 1.97                    |
| [39]   | Metasurface      | 6.6/11.7               | 0.0646/0.0268         | 5.2/5.8                 |

## 5. CONCLUSION

Use of EBG, FSS, and metamaterials are all effective techniques to reduce SAR in antennas, and their effectiveness depends on the specific design requirements and operating frequencies of the antenna. It is important to carefully evaluate the antenna's design requirements and operating frequencies to deter- mine which technique is better suited for a specific application. EBG structures use periodic arrays of conductive or dielectric elements to block or reflect electromagnetic waves and are effective in reducing SAR in antennas at lower frequencies and can also improve the antenna's directivity, whereas FSS struc- tures are thin periodic sheets of conductive or dielectric material that filter out specific frequencies and reduce the radiated energy in certain frequency bands. FSS structures are effective in reducing SAR in antennas at higher frequencies and can also improve the antenna's radiation efficiency. However, FSS structures can be challenging to design and can reduce the antenna's bandwidth. Meta-materials can be used to reduce SAR by controlling the radiation pattern of the antenna and reducing the energy radiated towards the human body. Meta-materials are also effective for a wide range of frequencies and can improve the antenna's performance.

## REFERENCES

1.  H. Li, J. Du, X.-X. Yang, and S. S. C. Gao, "Low-profile all-textile multiband microstrip circular patch antenna for WBAN applications," IEEE Antennas Wireless Propag. Lett., early access, Jan.27, 2022.
2.  S.H.Li and J.S.Li,"Smart patch wearable antenna on Jeans textile for body wireless communication," 2018 12th International Symposium on Antennas, Propagation and EM Theory (ISAPE), Hangzhou, China, 2018.

3. A. Anbalagan, E. F. Sundarsingh, V. S. Ramalingam, A. Samdaria, D. B. Gurion, and K. Balamurugan, "Realization and analysis of a novel low-profile embroidered textile antenna for real-time pulse monitoring," IETE J. Res., pp. 1–8, Jul. 2020.

4. Yao, L.; Li, E.; Yan, J.; Shan, Z.; Ruan, X.; Shen, Z.; Ren, Y.; Yang, J.,"Miniaturization and Electromagnetic Reliability of Wearable Textile Antennas.", Electronics 2021.

5. X. Lin, Y. Chen, Z. Gong, B. -C. Seet, L. Huang and Y. Lu, "Ultrawideband Textile Antenna for Wearable Microwave Medical Imaging Applications," in IEEE Transactions on Antennas and Propagation, vol. 68, no. 6, pp. 4238-4249, June 2020.

6. Yadav, A.; Singh, V.K.; Yadav, P.; Beliya, A.K.; Bhoi, A.K.; Barsocchi, P. Design of Circularly Polarized Triple-Band Wearable Textile Antenna with Safe Low SAR for Human Health. Electronics 2020, 9, 1366.

7. P. Schilingovski, V. Vulfin, S. Sayfan-Altman and R. Shavit, "Wearable antennas design for wireless communication," 2017 IEEE International Conference on Microwaves, Antennas, Communications and Electronic Systems (COMCAS), Tel-Aviv, Israel, 2017.

8. P. M. Potey and K. Tuckley, "Design of wearable textile antenna for low back radiation," J. Electromagn. Waves Appl., vol. 34, no. 2, pp. 235–245, Dec. 2019.

9. M. M. H. Mahfuz, M. R. Islam, N. Sakib, M. H. Habaebi, R. Raad and M. A. Tayab Sakib, "Design of Wearable Textile Patch Antenna Using C-Shape Etching Slot for Wi-MAX and 5G Lower Band Applications," 2021 8th International Conference on Computer and Communication Engineering (ICCCE), Kuala Lumpur, Malaysia, 2021.

10. I. Gil and R. Fernández-García, "SAR impact evaluation on jeans wearable antennas," 2017 11th European Conference on Antennas and Propagation (EUCAP), Paris, France, 2017.

11. Shah, A. and Patel, P., "E-textile slot antenna with spurious mode suppression and low SAR for medical wearable applications", Journal of Electromagnetic Waves and Applications, vol. 35, no.16, pp. 2224–2238, 2021.

12. Shirvani, P. Khajeh-Khalili, F. and M. H. Neshati, "Design investigation of a dual-band wearable antenna for tele-monitoring applications," AEU Int. J. Electronics and Communication, vol. 138, Aug. 2021.

13. Bo Yin Jing Gu Xingxing Feng Bin Wang Youhai Yu Wei Ruan , "A Low SAR Value Wearable Antenna for Wireless Body Area Network Based on AMC Structure," Progress In Electromagnetics Research C, Vol. 95, 119-129, 2019.

14. Gao, G, Meng, H, Geng, W, Zhang, B, Dou, Z, Hu, B. ,"Design of a wide bandwidth and high gain wearable antenna based on nonuniform metasurface",. Microw Opt TechnolLett. 2021.

15. V. R. Keshwani and S. S. Rathod, "Assessment of SAR reduction in Wearable Textile Antenna," 2021 International Conference on Communication information and Computing Technology(ICCICT), Mumbai, India, 2021.

16. Florence, Esther S., Malathi Kanagasabai, Gulam Nabi and Mohammed Gulam Nabi Alsath. "An Investigation of a Wearable Antenna Using Human Body Modelling." Appl. Comput. Electromagn. Soc. J., vol. 29, no. 10, pp. 777–783, Oct. 2014.

17. S. Zhu and R. Langley, "Dual-Band Wearable Textile Antenna on an EBG Substrate," in IEEE Transactions on Antennas and Propagation, vol. 57, no. 4, pp. 926-935, April 2009.

18. S. Velan et al., "Dual-Band EBG Integrated Monopole Antenna Deploying Fractal Geometry for Wearable Applications," in IEEE Antennas and Wireless Propagation Letters, vol. 14, pp. 249-252, 2015.

19. Mahmood, S.N.; Ishak, A.J.; Saeidi, T.; Soh, A.C.; Jalal, A.; Imran, M.A.; Abbasi, Q.H. Full Ground Ultra-Wideband Wearable Textile Antenna for Breast Cancer and Wireless Body Area Network Applications. Micromachines 2021.

20. A. Y. I. Ashyap et al., "Highly Efficient Wearable CPW Antenna Enabled by EBG-FSS Structure for Medical Body Area Network Applications," in IEEE Access, vol. 6, pp. 77529-77541, 2018.

21. Ali, Usman; Ullah, Sadiq; Khan, Jalal; Shafi, Muhammad; Kamal, Babar; Basir, Abdul," Design and SAR analysis of wearable antenna on various parts of human body, using conventional and artificial ground planes",. Loughborough University. Journal contribution (2017)

22. K. N. Paracha et al., "A Low Profile, Dual-band, Dual Polarized Antenna for Indoor/Outdoor Wearable Application," in IEEE Access, vol. 7, pp. 33277-33288, 2019.

23. S. I. Kwak, D. -U. Sim and J. H. Kwon, "Design of Optimized Multilayer PIFA With the EBG Structure for SAR Reduction in Mobile Applications," in IEEE Transactions on Electromagnetic Compatibility, vol. 53, no. 2, pp. 325-331, May 2011.

24. A. Y. I. Ashyap, Z. Z. Abidin, S. H. Dahlan, H. A. Majid and F. C. Seman, "A Compact Wearable Antenna Using EBG for Smart-Watch Applications," 2018 Asia-Pacific Microwave Conference (APMC), Kyoto, Japan, 2018.

25. A. Y. I. Ashap, Z. Z. Abidin, S. H. Dahlan, H. A. Majid, S. K. Yee, G. Saleh, and N. A. Malek, "Flexible wearable antenna on electromagnetic band gap using PDMS substrate," TELKOMNIKA (Telecommun. Comput. Electron. Control), vol. 15, no. 3, p. 1454, Sep. 2017.

26. A. Y. I. Ashyap, Z. Z. Abidin, S. H. Dahlan, H. A. Majid, M. R. Kamarudin and R. A. Abd- Alhameed, "Robust low-profile electromagnetic band-gap-based on textile wearable antennas for medical application," 2017 International Workshop on Antenna Technology: Small Antennas, Innovative Structures, and Applications (iWAT), Athens, Greece, 2017.

27. M. I. Ahmed, E. A. Abdallah, and H. M. Elhennawy, "SAR investigation of novel wearable Reduced- coupling microstrip antenna array," Int. J. Eng., vol. 15, no. 3, pp. 78–86, 2015.

28. Gao, G, Hu, B, Wang, S, Yang, C. Wearable planar inverted-F antenna with stable characteristicand low specific absorption rate. Microw Opt Technol Lett. 2018.

29. M. T. Islam, N. Misran, T. S. Ling and M. R. I. Faruque, "Reduction of specific absorption rate (SAR) in the human head with materials and metamaterial," 2009 International Conference on Electrical Engineering and Informatics, Bangi, Malaysia, 2009.

30. Mohammad Tariqul Islam, Mohammad Rashed, Iqbal Faruque, Norbahiah Misran, "Electromag- netic (EM) absorption rate reduction of helix antenna with shielding material for mobile phone application", Australian Journal of Basic and Applied Sciences 2011.

31. S. i. Kwak, D. -U. Sim, J. H. Kwon and Y. J. Yoon, "Design of PIFA With Metamaterials for Body-SAR Reduction in Wearable Applications," in IEEE Transactions on Electromagnetic Compatibility, vol. 59, no. 1, pp. 297-300, Feb. 2017.

32. M. Wang et al., "Investigation of SAR Reduction Using Flexible Antenna With Metamaterial Structure in Wireless Body Area Network," in IEEE Transactions on Antennas and Propagation, vol. 66, no. 6, pp. 3076-3086, June 2018.

33. Bashyam Sugumaran, Ramachandran Balasubramanian, Sandeep Kumar Palaniswamy ," Reduced specific absorption rate compact flexible monopole antenna system for smart wearable wireless communications," Engineering Science and Technology, an International Journal, 2021.

34. A. Y. I. Ashyap et al., "Fully Fabric High Impedance Surface-Enabled Antenna for Wearable Medical Applications," in IEEE Access, vol. 9, pp. 6948-6960, 2021.

35. S. M. Saeed, C. A. Balanis, C. R. Birtcher, A. C. Durgun and H. N. Shaman, "Wearable Flexible Reconfigurable Antenna Integrated With Artificial Magnetic Conductor," in IEEE Antennas and Wireless Propagation Letters, vol. 16, pp. 2396-2399, 2017.

36. K. Agarwal, Y. -X. Guo and B. Salam, "Wearable AMC Backed Near-Endfire Antenna for On- Body Communications on Latex Substrate," in IEEE Transactions on Components, Packaging andManufacturing Technology, vol. 6, no. 3, pp. 346-358, March 2016.

37. Yalduz, H., Tabaru, T.E., Kılıç, V.T., and Turkmen, M. (2020). Design and analysis of low profile and low SAR full-textile UWB wearable antenna with metamaterial for WBAN applications. AEU
- International Journal of Electronics and Communications.

38. S. i. Kwak, D. -U. Sim, J. H. Kwon and Y. J. Yoon, "Design of PIFA With Metamaterials for Body-SAR Reduction in Wearable Applications," in IEEE Transactions on Electromagnetic Compatibility, vol. 59, no. 1, pp. 297-300, Feb. 2017.

39. He-Lin Yang, Wang Yao, Yuanyuan Yi, Xiaojun Huang, Song Wu, and Boxun Xiao , "A Dual-Band Low-Profile Metasurface-Enabled Wearable Antenna for WLAN Devices," Progress In Electromagnetics Research C, Vol. 61, 115-125, 2016.